\documentclass[twocolumn,tighten]{aastex6}

\usepackage{graphicx}	
\usepackage{amsmath}	
\usepackage{amssymb}	
\usepackage[hyphenbreaks]{breakurl}
\usepackage{upgreek}

\newcommand\swiftj{{\rm~Swift~J1658.2--4242}}

\begin{document}
\title{Spectral and Timing Properties of the Galactic X-ray transient Swift~J1658.2--4242 using {\it Astrosat} Observations}
\shorttitle{Spectral and Timing Properties of Swift~J1658.2--4242}

\author{V. Jithesh$^1$, Bari Maqbool$^1$, Ranjeev Misra$^1$, Athul. R. T$^2$, Gitika Mall$^3$ and Marykutty James$^4$ }
\affil{$^1$ Inter-University Centre for Astronomy and Astrophysics (IUCAA), Post Bag No. 4, Ganeshkhind, Pune 411007, India; vjithesh@iucaa.in\\
$^2$ Department of Physics, Cochin University of Science \& Technology, Kochi 682022, India \\
$^3$ Department of Physics, Banaras Hindu University, Varanasi-221005, India \\ 
$^4$ Department of Physics, St. Thomas college, Ranni, Pathanamthitta 689673, Kerala, India}
\shortauthors{Jithesh et al}

\begin{abstract}
We present the X-ray timing and spectral analysis of the new Galactic X-ray transient Swift J1658.2-4242 observed with LAXPC and SXT instruments onboard {\it Astrosat}. We detect prominent C-type quasi-periodic oscillations (QPOs) of frequencies varying from $\sim 1.5$ Hz to $\sim 6.6$ Hz along with distinct 2nd harmonics and sub-harmonics. The QPO detected at $\sim 1.56$ Hz drifts to a higher centroid frequency of $\sim 1.74$ in the course of the observation, while the QPO detected at $\sim 6.6$\,Hz disappeared during hard flarings. The fractional rms at the QPO and the sub-harmonic frequencies increases with photon energy, while at the 2nd harmonic frequencies the rms seems to be constant. In addition, we have observed soft time lag at QPO and sub-harmonic frequencies up to a time scale of $\sim 35$ ms, however, at the 2nd harmonic frequencies there is weak/zero time lag.  We attempt spectral modeling of the broadband data in the 0.7--25 keV band using the doubly absorbed disk plus thermal Comptonization model. Based on the spectral and timing properties, we identified the source to be in the hard intermediate state of black hole X-ray binaries. To quantitatively fit the energy and frequency-dependent fractional rms and time lag, we use a single zone fluctuation propagation model and discuss our results in the context of that model.    
\end{abstract}


\section{Introduction} 
\label{sec:intro}
Black hole transients (BHTs) spend most of their lives in quiescence and are primarily discovered when they enter into outbursts characterized by abrupt changes in their X-ray luminosity by several orders of magnitudes. During typical outbursts, BHTs undergo a transition from the low hard state (LHS) to the high soft state (HSS) through short-lived hard and soft intermediate states (HIMS, SIMS), then again back to the LHS via the intermediate states \citep{Rem06, Bel10, Bel11}. In the LHS, the source is characterized by a hard spectrum with high ($\sim 30\%$) fractional rms variability \citep{Bel05} and sometimes show low-frequency quasi-periodic oscillations (LFQPOs) in the power density spectrum (PDS). A soft thermal component modeled with multi-color disk blackbody component dominates in the HSS, while the fractional rms reduces down to a few percents and PDS shows weak power law noise. 

The intermediate states identified in the black hole systems are rather complex and the observed behaviors are difficult to interpret. In these states, the energy spectrum contains both disk and the power law components, while the PDS mainly contains LFQPOs with centroid frequency ranging from a few mHz to $\sim 30$\,Hz. LFQPOs are identified in several sources and they are classified into type A, B, and C \citep{Rem02, Cas05}. Type C QPOs are mainly identified in the HIMS, while type A and B QPOs are detected in the SIMS \citep{Wij99, Cas05, Mot11}. Although the origin of the LFQPOs is still under debate, several models have been proposed to explain the origin and evolution of LFQPOs in X-ray binaries. The proposed models are generally based on the two different mechanisms: instabilities \citep[e.g.][]{Tag99, Tit99, Tit04, Cab10} and geometrical effects \citep{Ste98, Ste99, Ing09, Ing11}. LFQPOs are widely detected in BHTs and their detailed studies provide essential information regarding the accretion flow around the black hole and geometry of the system. In addition, it is important to examine the energy-dependent properties of QPOs such as fractional rms and time lag, which also provide the link between the spectral and timing variability properties of BHTs. 

\swiftj{} is a new Galactic X-ray transient source discovered by the BAT instrument onboard {\it Swift} on 2018 February 16 \citep{Bar18}. The IBIS/ISGRI instruments onboard {\it INTEGRAL} detected the source to be in the hard state during their observations of the Galactic center field performed from 2018 February 13 to 16 \citep{Gre18}. The reported position of the source from the {\it Swift} XRT observation is R.A. = 16:58:12.58, Decl. = -42:41:55.7 (equinox J2000.0) with an uncertainty of $\sim 4$ arcsec \citep{Dav18}. Radio observation with the {\it Australia Telescope Compact Array} ({\it ATCA}) identified a radio source at a position consistent with the {\it Swift} XRT position and the radio emission was consistent with a flat radio spectrum from a compact jet implying that the source is a black hole X-ray binary (BHXRB) at distance $> 3$ kpc \citep{Rus18}. In the positional uncertainty of {\it ATCA}, an optical source has been observed in the archival imaging observation and the optical spectroscopy observation with SOAR telescope suggest this source is a normal, mid to late-type K star with no signatures of accretion \citep{Bah18}. The {\it Swift} XRT observation taken with Window Time (WT) mode reports the presence of a peak at 0.115\,Hz in the power spectrum \citep{Ken18}, which was explained as a coherent period of the source rather than a QPO, and the authors suggested that the system may be a Be/X-ray binary with a neutron star compact object.    

\citet{Xu18} presented X-ray timing and spectral studies of this source using simultaneous {\it NuSTAR} and {\it Swift} observations. The broadband spectral modeling with relativistic reflection models suggested that the source is a BHXRB system which is observed in the bright hard state with a photon index of $\sim 1.6$ and coronal temperature of $\rm kT_{e} \sim 22$ keV. From the relativistic disk reflection features, they suggest a highly spinning black hole for the system with a spin value $> 0.96$ and the inner accretion disk is viewed at a high inclination angle ($\sim 64^\circ$). In addition, a low frequency (type-C) QPO has been observed in the power spectrum with a shift in the QPO frequency from $\sim 0.14$ to $\sim 0.21$\,Hz during the single {\it NuSTAR} observation. 

Recently, \citet{Xu19} studied the source in the intermediate state using the {\it NuSTAR}, {\it XMM-Newton} and {\it Swift} observations. During these observations, the source intensity decreased rapidly by 45\%, after which a transient QPO at 6-7\,Hz was detected in the low flux state. The authors discussed these observed properties by invoking the accretion disk instability scenario. In addition, the relativistic disk reflection component detected in the bright hard state became weaker and not significantly detected in these observations. Further, the rapid variation of X-ray flux leads to only subtle changes in the shape of the broadband X-ray spectrum, even though the disk temperature decreases by $\sim 15\%$. 

These observed properties of the source can further be investigated using {\it Astrosat} data. \swiftj{} was observed by {\it Astrosat} on 2018 February 20 and the preliminary analysis with Large Area X-ray Proportional Counter instrument on-board {\it Astrosat} detected a sharp QPO at 1.6\,Hz with r.m.s value of $16\%$. The observed centroid frequency increased from $\sim 1.6$ to 2\,Hz during the 20 ks observation \citep{Ber18}.     
  
In this work, we studied the X-ray timing and spectral characteristics of \swiftj{} using {\it Astrosat} observations. {\it Astrosat} \citep{Sin14, Agr17} is India's first multi-wavelength astronomical observatory, which contains five instruments on-board: Soft X-ray Telescope (SXT), Large Area X-ray Proportional Counter (LAXPC), Cadmium Zinc Telluride Imager (CZTI), Scanning Sky Monitor (SSM), an Ultra Violet Imaging Telescope (UVIT). We analyze the simultaneous observations of \swiftj{} using the SXT and LAXPC instruments on board {\it Astrosat}. Section \ref{sec:obs} describes the observations used in the work and the data reduction techniques. The analysis and results are presented in \S \ref{sec:resu}. The main results are summarized and discussed in \S \ref{sec:discu}.

\section{Observations and Data Reduction}
\label{sec:obs}

\begin{figure}
\begin{center}

\includegraphics[width=8.5cm,angle=0]{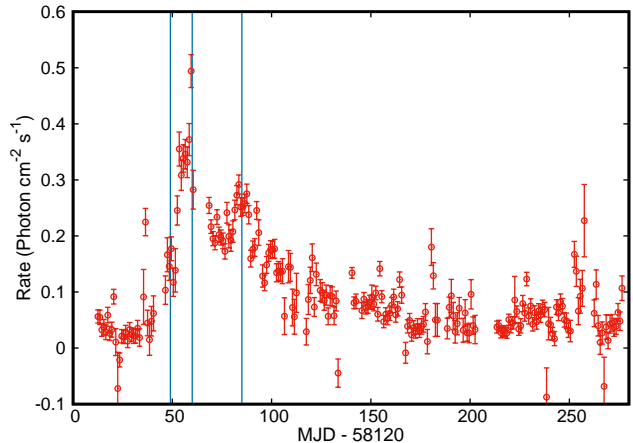}
\caption{The 1-day binned {\it MAXI} light curve of \swiftj{} in the 2--20 keV energy band over the period 2018 January 14 to October 9. The vertical blue lines represent three {\it Astrosat} observations considered in this work.}
\label{maxi-lc}
\end{center}
\end{figure}

We used three publicly available {\it Astrosat} Target of Opportunity (ToO) observations of the recently discovered X-ray transient \swiftj{} in the Galactic plane. The first observation (O1; ObsID: T02\_004T01\_9000001910) was conducted during 2018 February 20--21 for an effective exposure of $\sim 20$ ks. In the second observation (O2), the source was observed for an effective exposure time $\sim 30$ ks during 2018 March 03--04 (ObsID: T02\_011T01\_9000001940) and the third observation (O3) was performed on 2018 March 28--29 (ObsID: T02\_020T01\_9000001990) with an effective exposure of 30 ks. The observations considered in the analysis are marked on the {\it MAXI} light curve shown in Figure \ref{maxi-lc}.   

\subsection{Soft X-ray Telescope}
SXT \citep{Sin16, Sin17} is a focusing X-ray telescope with CCD camera operates in the 0.3--8 keV energy band. We processed the Level-1 Photon Counting (PC) mode data using the SXT pipeline software\footnote{\url{http://www.tifr.res.in/~astrosat\_sxt/sxtpipeline.html}} (version: AS1SXTLevel2-1.4a) and obtained the cleaned Level-2 event files from all orbits. The SXT event merger script\footnote{\url{http://www.tifr.res.in/~astrosat\_sxt/dataanalysis.html}\label{link2}} was used to merge different orbits data and obtained an exposure corrected, merged cleaned event file. We extracted the source spectrum from the merged event file using a circular region of radius 16 arcmin with the standard tools available in XSELECT V2.4d. The 16 arcmin extraction radius includes $\sim 97\%$ of the total photons from the source. The SXT off-axis auxiliary response file (ARF) is generated by the latest {\tt sxtARFModule} tool$^2$ using the latest on-axis ARF (version 20190608) provided by the SXT instrument team. We used the blank sky SXT spectrum, provided by the instrument team, as the background spectrum. The command {\tt gain} is used to modify the gain of the response file when fitting the SXT spectrum. The slope is fixed at unity and the offset is free to vary. In this work, we used the SXT spectrum in the 0.7--7 keV energy band for the spectral model fitting. 

\subsection{Large Area X-ray Proportional Counter}
LAXPC consists of three identical proportional counters ({\tt LAXPC10, LAXPC20} and {\tt LAXPC30}) operates in the 3--80 keV energy band with an absolute time resolution of $10\,\rm \mu s$  \citep{Yad16, Yad16a, Ant17, Agr17a}. All observations of the source are taken in Event Analysis (EA) mode. We processed the data using the LAXPC software\footnote{\url{http://astrosat-ssc.iucaa.in/?q=laxpcData}} (LaxpcSoft; version as of 2018 May 19) and generated the Level 2 event file. The light curve and energy spectrum were extracted from the Level 2 event file using the standard tasks available in LaxpcSoft \footnote{\url{http://www.tifr.res.in/~astrosat\_laxpc/LaxpcSoft.html}}. {\tt LAXPC30} detector was affected by the gas leakage and switched off on 2018 March 8 due to abnormal gain changes, hence we do not include the {\tt LAXPC30} detector in our analysis. In the third observation (O3), the {\tt LAXPC10} detector was operating at low gain,  thus we continued our analysis by considering only {\tt LAXPC20}. Since the background dominates above 25 keV, we used 4--25 keV energy band for the LAXPC spectrum. 

\begin{figure}
\begin{center}

\includegraphics[width=8.5cm,angle=0]{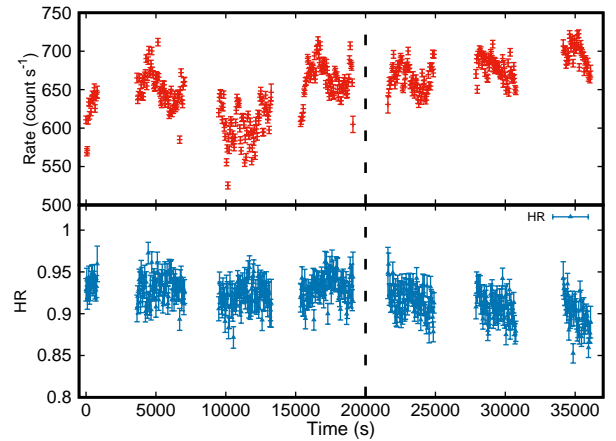}

\caption{The 3--50 keV energy band light curve from the {\tt LAXPC10} and {\tt LAXPC20} detectors (top panel) and the hardness ratio (bottom panel) of \swiftj{} from the first observation. The light curve is background subtracted and binned with 50\,s. The black dashed vertical line separates the time duration where the two QPOs ($\sim 1.56$ and $\sim 1.74$\,Hz) detected.} 
\label{lc_data1}
\end{center}
\end{figure}

\begin{figure}
\begin{center}

\includegraphics[width=6.2cm,angle=-90]{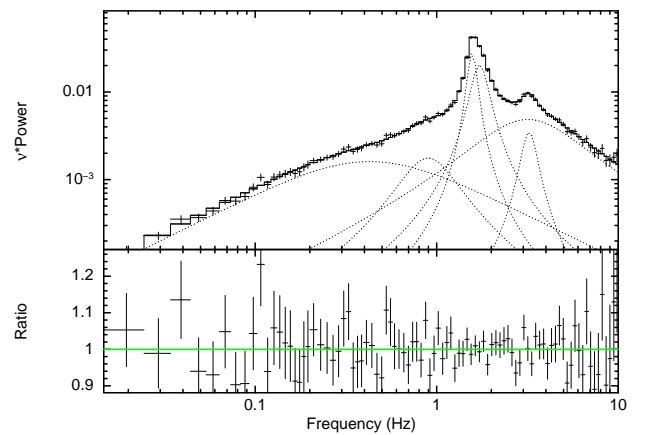}
\caption{Power density spectrum of \swiftj{} from the first observation in the 3--15 keV fitted with six Lorentzians.}
\label{pds_data1}
\end{center}
\end{figure}

\section{Analysis and Results}
\label{sec:resu}

\subsection{Timing Analysis}

\subsubsection{Observation 1}

We extracted the background subtracted light curve of the source from the LAXPC instrument in the 3--50 keV energy band, which is shown in the top panel of Figure \ref{lc_data1}. We used the {\tt LAXPC10} and {\tt LAXPC20} detectors for the light curve extraction with a time bin size of 50\,s. The source intensity appears to be constant with an average count rate of $\sim 654~\rm count~s^{-1}$, with a marginal increase in the last part of the observation. The hardness ratio (HR) is calculated using the count rate in the 3--7 and 7--16 keV bands, and plotted in the bottom panel of Figure \ref{lc_data1}. From the plot, it is clear that the HR is in the range of $\sim 0.85-0.97$ and we do not see any particular trend in the observation. To check the presence of QPOs in the light curve, we extracted the PDS in the 3--15 keV, 15--30 keV and 30--50 keV energy bands. The extracted PDS in the 3--15 keV band is fitted with six Lorentzians and shown in Figure \ref{pds_data1}. A QPO is detected at $\sim 1.6$ Hz in the PDS with a 2nd harmonic at $\sim 3.2$ Hz. However, the detected QPO feature cannot be fitted with a single Lorentzian. Thus, we attempted to fit the broad feature with two Lorentzians with centroid frequencies of $\sim 1.57$ and $\sim 1.71$ Hz. Moreover, it is noticed that the 2nd harmonic feature observed in the 3-15 keV band is weak and not clearly detected in the 15--30 keV and 30--50 keV energy bands.  

\begin{figure}
\includegraphics[width=9.0cm,angle=0]{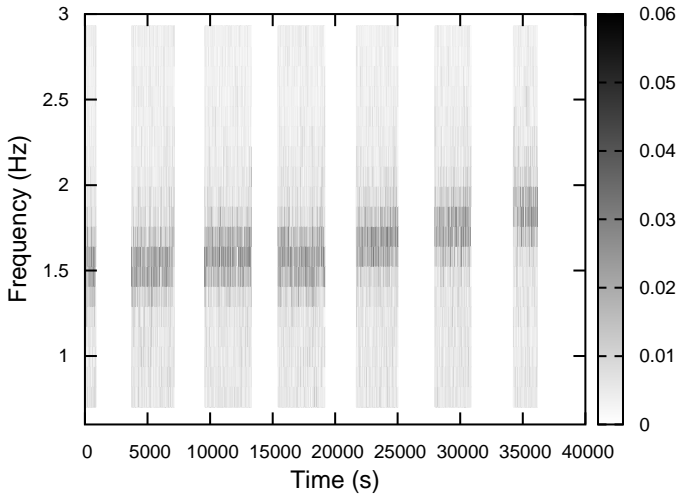}
\caption{The dynamic power spectrum of \swiftj{} from the first observation, which clearly shows the change in the QPO frequency from $\sim 1.56$ to $\sim 1.74$ Hz within the observation.}
\label{dps_data1}
\end{figure}

\begin{figure*}
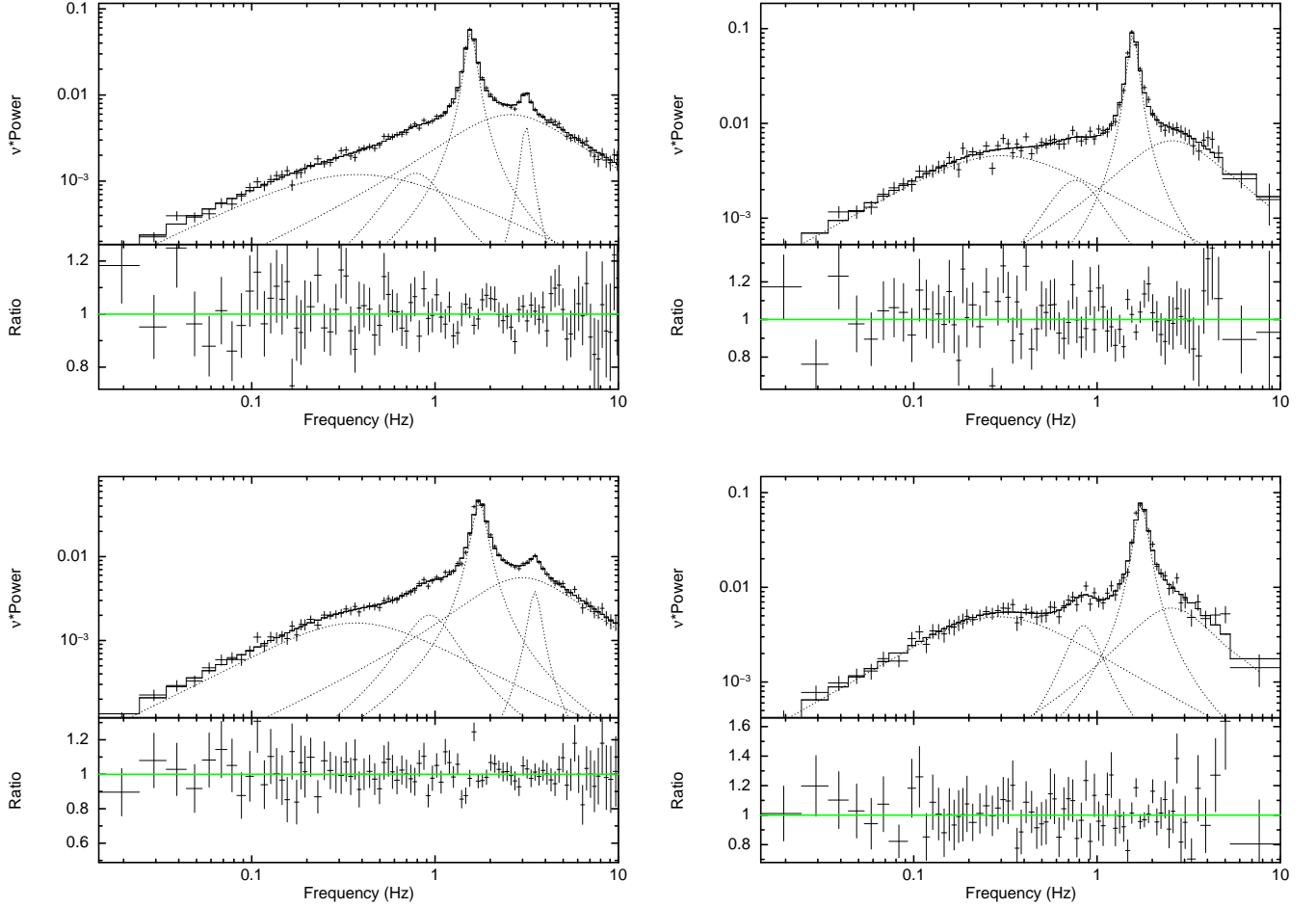


 \includegraphics[width=6.5cm,angle=-90]{f5a.ps}
 \includegraphics[width=6.5cm,angle=-90]{f5b.ps}

 \includegraphics[width=6.5cm,angle=-90]{f5c.ps}
 \includegraphics[width=6.5cm,angle=-90]{f5d.ps}

\caption{Power density spectra from the two time segments in the first observation fitted with five Lorentzians. The top and bottom panels for the first and second time segments, respectively. The PDS in the 3--15 keV and 15--30 keV energy range are depicted in the left and right panels, respectively.}
\label{seg_data1}
\end{figure*}

Since the source exhibited an increasing trend in the centroid frequency within the {\it NuSTAR} observation \citep{Xu18}, we tested the time-dependent behavior of the QPO by extracting the dynamic power spectrum (DPS) using the standard LAXPC analysis tools. We used the frequency interval of 0.1--30\,Hz to compute the DPS. It is clear from the DPS, which is shown in Figure \ref{dps_data1}, that the QPO changes its centroid frequency from $\sim 1.56$ to $\sim 1.74$ Hz during the observation. Thus, we split the light curve into two parts with time segments below 20 ks and above 20 ks in the light curve (see also Figure \ref{lc_data1}). To confirm the QPO frequency change, we again extracted the PDS with these time segments, which are shown in Figure \ref{seg_data1}. The PDS from the two time segments confirmed the change in the QPO frequency, and the 2nd harmonic feature is strongly detected in both time segments only in the 3--15 keV band. 

\begin{figure*}

\includegraphics[width=8.9cm,height=7.0cm]{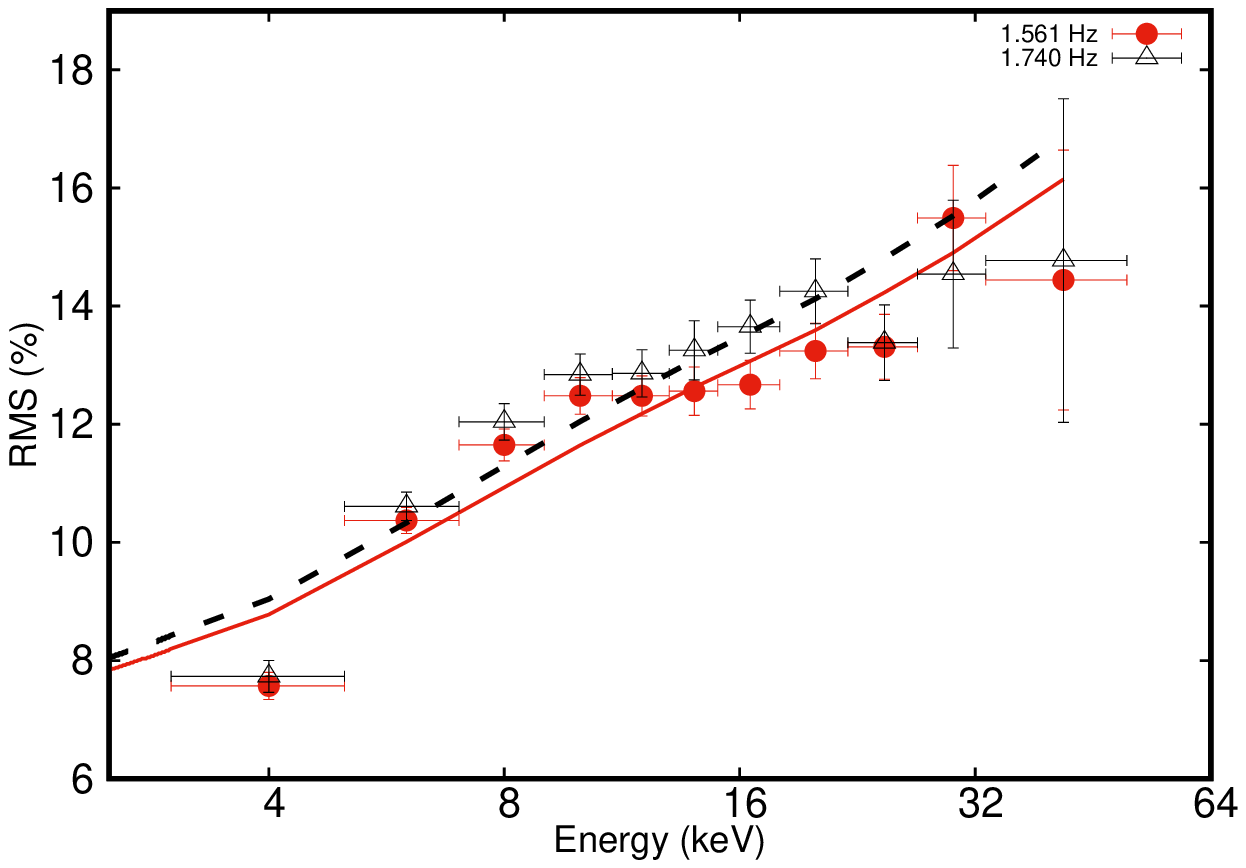}
\includegraphics[width=8.9cm,height=7.0cm]{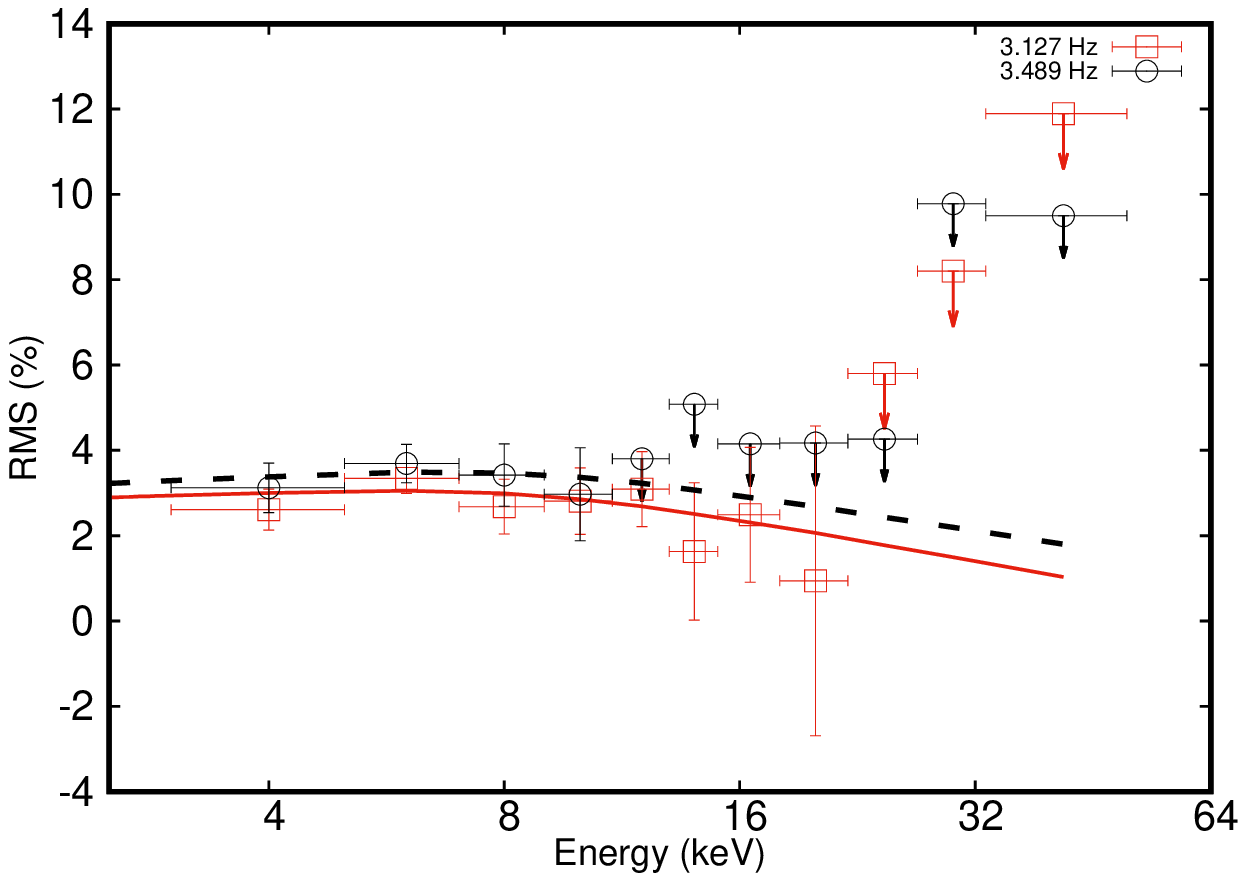}

\includegraphics[width=8.9cm,height=7.0cm]{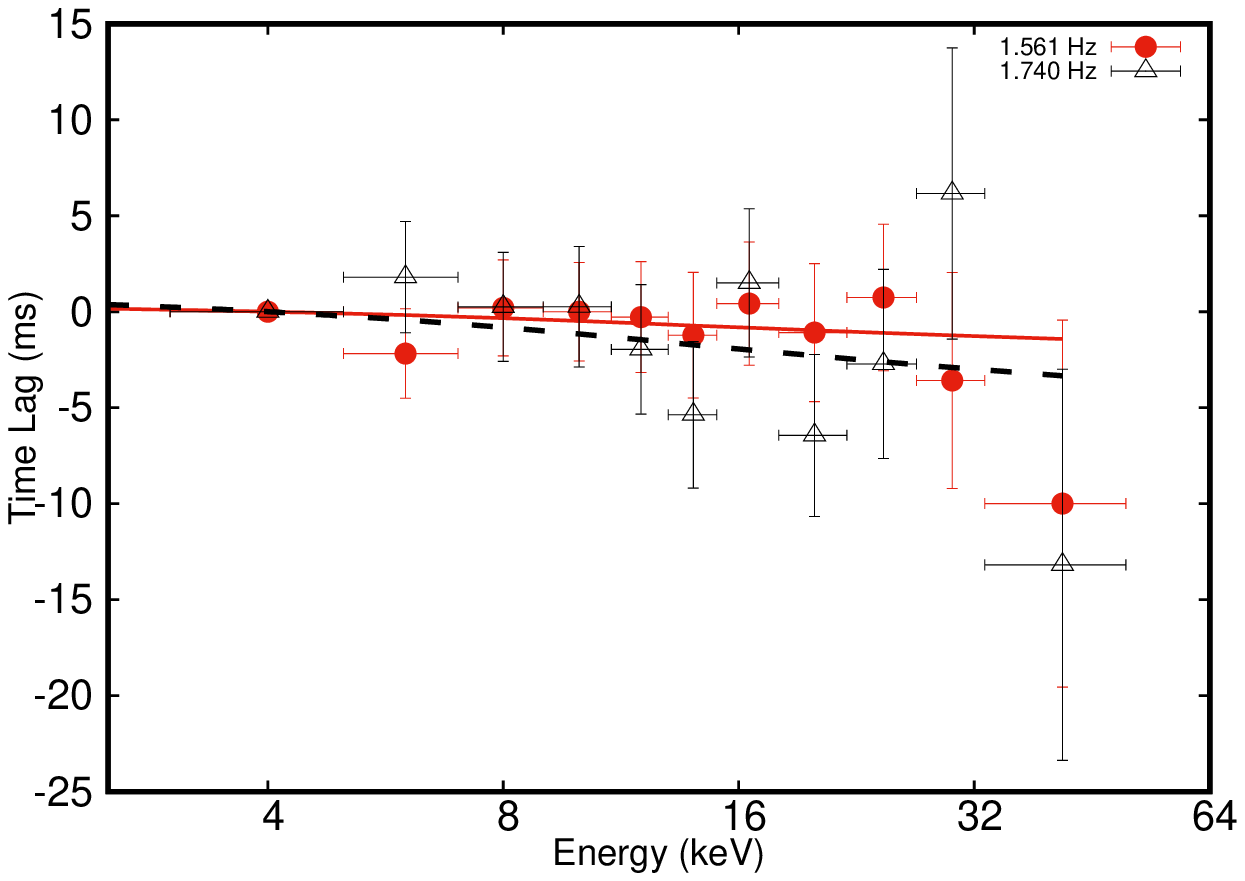}
\includegraphics[width=8.9cm,height=7.0cm]{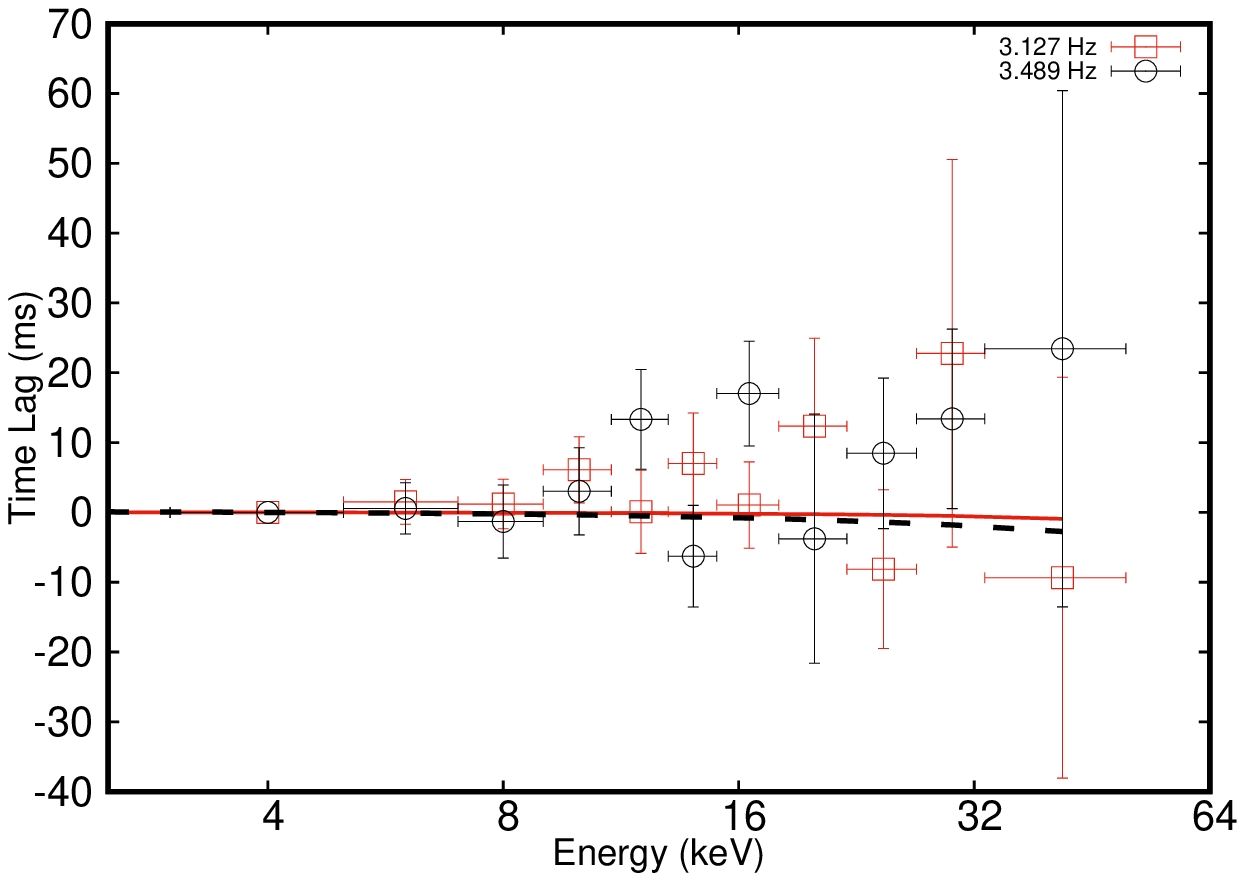}

\caption{Fractional rms (top panels) and time lag (bottom panels) as a function of photon energy. The left panels are for the fundamentals, while the right panels are for the 2nd harmonic. The red filled circle and black open triangle represent $\sim 1.56$ Hz (first segment) and $\sim 1.74$ Hz (second segment) QPOs, respectively. The red open square and the black open circle represent the 2nd harmonic frequencies at $\sim 3.13$ Hz (first segment) and $\sim 3.49$ Hz (second segment), respectively. The red solid line and black dotted line represent the model fit derived from the fluctuation propagation model (see text for details).}
\label{data1_seg1_2_rm_lag}
\end{figure*}

We determine the fractional rms as a function of energy for the QPOs using several energy bands. We extracted the PDS for these energy bands and fitted the PDS with five Lorentzians. In the model fit, we fix the centroid frequency and width of all Lorentzians to the values obtained from the fitting the 3--15 keV PDS and vary the normalization. The fractional rms is estimated by taking the square root of the normalization of the Lorentzian representing the QPO feature. We used the same energy bands to estimate the time lag by following the method in \citet{Now99}. The calculated fractional rms and time lag for the two QPOs at $\sim 1.56$ and $\sim 1.74$ Hz by considering the respective time segments are plotted in Figure \ref{data1_seg1_2_rm_lag}. For the time lag estimation, the frequency resolution was taken to be $\sigma/3$, where $\sigma$ is the width of the QPO. The strength of the QPOs showed a clear variation with photon energy in both time segments as shown in the top left panel of Figure \ref{data1_seg1_2_rm_lag}. For the QPO at $\sim 1.56$ Hz, the fractional rms rises from $\sim 7$ percent and reaches a maximum of $\sim 15$ percent at the 27--33 keV energy band, but marginally decreases in the 33--50 keV band. For the QPO at $\sim 1.74$ Hz observed in the second time segment, the fractional rms gradually increases from $\sim 8$ percent to reach a maximum value of $\sim 15$ percent in the highest (33--50 keV) energy band. The fractional rms of the 2nd harmonics, detected at $\sim 3.13$ and $\sim 3.49$ Hz, seems to be constant in the different energy bands (see top right panel of Figure \ref{data1_seg1_2_rm_lag}). The time lags for the QPOs and 2nd harmonics are not well constrained. 

\begin{table*}
\tablecolumns{7}
\setlength{\tabcolsep}{8.0pt}
\tablewidth{320pt}
	\caption{QPO, 2nd harmonic and sub-harmonic Parameters of \swiftj{} in different observations}
 	\begin{tabular}{@{}ccccccccc@{}}
	\hline
	\hline
\colhead{Obs} & \colhead{Seg} & \colhead{QPO} & \colhead{QPO} & \colhead{2nd Harmonic} & \colhead{{Width}} & \colhead{Sub-harmonic} & \colhead{{Width}} & \colhead{$\rm \chi^2/ d.o.f$} \\
 & & frequency (Hz) & width & frequency (Hz) & & frequency (Hz) &  & \\
\hline
O1 & -   & $1.566^{+0.011}_{-0.011}$ & $0.182^{+0.033}_{-0.038}$ & $3.211^{+0.044}_{-0.044}$ & $0.653^{+0.216}_{-0.172}$ & - & - & 67.1/73 \\  
      & -   & $1.712^{+0.042}_{-0.035}$ & $0.390^{+0.043}_{-0.052}$ &  &  &  &  &  \\
      & 1   & $1.561^{+0.004}_{-0.004}$ & $0.204^{+0.014}_{-0.014}$ & $3.127^{+0.031}_{-0.030}$ & $0.339^{+0.109}_{-0.099}$ & - & - & 82.4/76 \\
      & 2   & $1.740^{+0.006}_{-0.006}$ & $0.294^{+0.021}_{-0.021}$ & $3.489^{+0.050}_{-0.053}$ & $0.522^{+0.252}_{-0.188}$ & - & - & 88.4/ 76 \\

O2 & -  & $6.572^{+0.052}_{-0.052}$ & $2.032^{+0.189}_{-0.203}$ & - & - & $3.178^{+0.116}_{-0.134}$ & $1.847^{+0.509}_{-0.467}$ & 88.3/74 \\
   & 1  & $6.603^{+0.038}_{-0.038}$ & $1.963^{+0.119}_{-0.115}$ & - & - & $3.170^{+0.092}_{-0.104}$ & $1.773^{+0.336}_{-0.317}$ & 80.0/71 \\

O3 & -   & $3.975^{+0.040}_{-0.017}$ & $0.479^{+0.062}_{-0.075}$  & - & - & $2.287^{+0.106}_{-0.128}$ & $0.204^{+0.239}_{-0.201}$ & 91.5/47 \\
\hline
\end{tabular} 
\tablecomments {(1) Observation; (2) segment used; (3)-(4) Centroid frequency and width of QPOs (5)-(6) Centroid frequency and width of 2nd harmonic (7)-(8) Centroid frequency and width of sub-harmonic (9) $\chi^2$ statistics and degrees of freedom.}
\label{qpo_params}
\end{table*}

\subsubsection{Observation 2}

\begin{figure}

 \includegraphics[width=8.5cm,angle=0]{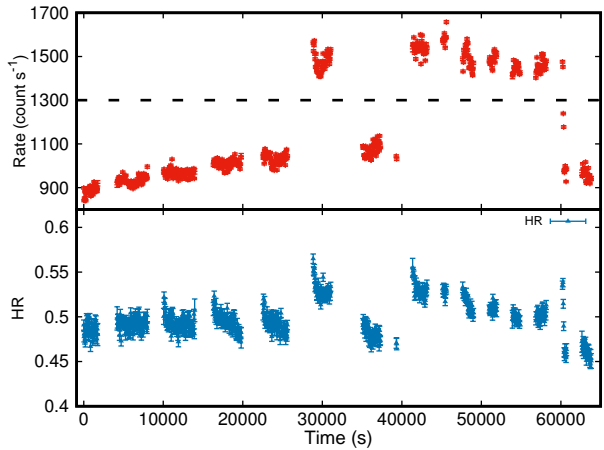}

\caption{The background subtracted light curve in the 3--50 keV energy (top panel) using the {\tt LAXPC10} and {\tt LAXPC20} detectors and hardness ratio (bottom panel) of the source from the second observation. We used a time bin size of 50\,s to extract the light curve. The dashed line in the top panel represents the two flux levels in the observation.}
\label{lc_data2}
\end{figure}

The extracted light curve in the 3-50 keV energy band from the second observation shows an interesting variation pattern in the intensity (see top panel of Figure \ref{lc_data2}). The source intensity increases gradually in the first part of the light curve followed by a drastic change in the intensity from $\sim 1000$ to $\sim 1500~\rm count~s^{-1}$ in a few kilo-second time. The intensity then drops suddenly to reach $\sim 1100~\rm count~s^{-1}$, which lasts for a few kilo-second after which the source again reaches the high intensity level ($\sim 1500~\rm count~s^{-1}$). The source intensity finally drops by a factor of $\sim 2$ in the last part of the observation. This variation in the intensity confirms the short term variability behavior of the source. We extracted HR from the 3--7 and 7--16 keV light curves as depicted in the bottom panel of Figure \ref{lc_data2}. We do see marginal HR variation corresponding to the variation pattern observed in the intensity. In addition, the source intensity in the second observation is higher than than the first observation by a factor of $\sim 3$.
 
\begin{figure}

\includegraphics[width=8.2cm,angle=0]{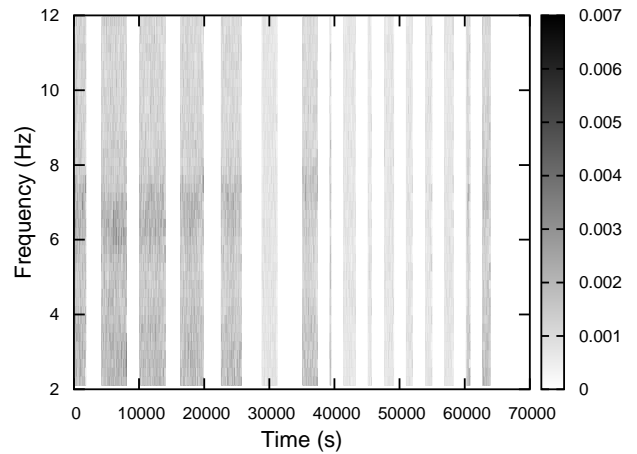}

\caption{The DPS of \swiftj{} from the second observation. It is clear that the $\sim 6.6$ Hz QPO is detected during the low intensity periods (see Figure \ref{lc_data2}).}
\label{dps_data2}
\end{figure}

\begin{figure*}
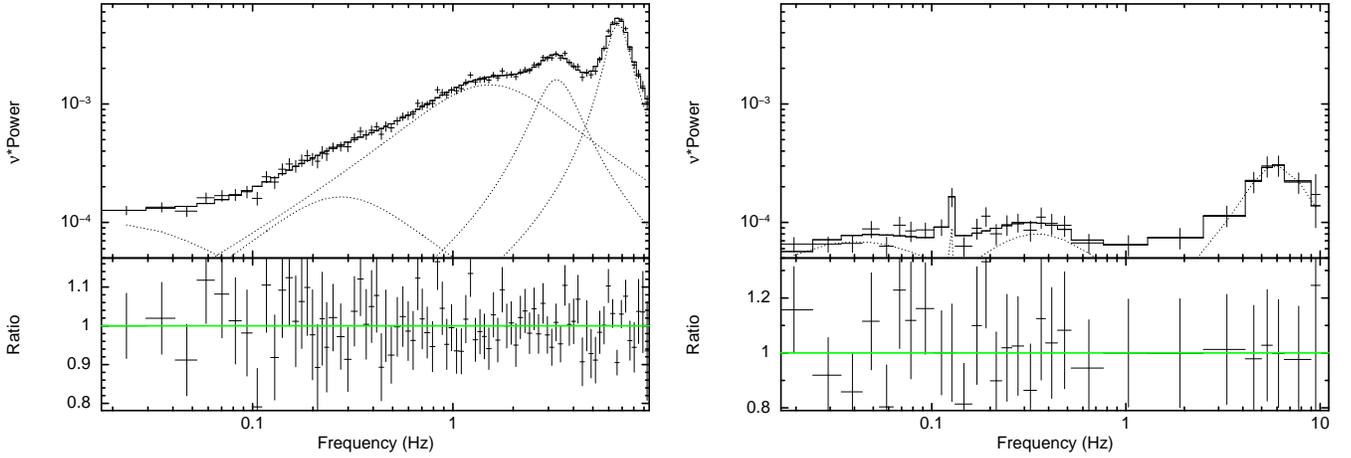


 \includegraphics[width=6.0cm,angle=-90]{f8a.ps}
 \includegraphics[width=6.0cm,angle=-90]{f8b.ps}

\caption{The PDS from the second observation in the 3--15 keV energy band. The left and right panels represent the low and high intensity level data, respectively.}
\label{pds_seg1_data2}
\end{figure*}

Since the source exhibits significant variability behavior, we extracted flux resolved light curve by choosing an intensity level of $1300~\rm count~s^{-1}$ (see also the top panel of Figure \ref{lc_data2}). Using the flux resolved light curve, we generated the PDS for the two intensity levels in the 3--15 and 15--30 keV bands. The PDS in low and high intensity levels are shown in the left and right panel of Figure \ref{pds_seg1_data2}. A QPO ($\sim 6.60$\,Hz) and a sub-harmonic ($\sim 3.17$\,Hz) is clearly detected in the PDS for the low flux level. However, it is interesting to note that the PDS in the high flux level is significantly changed compared to the PDS in low flux level and the QPO is weakly detected in the high flux level. We further computed the fractional rms of light curves for high and low flux levels from different energy bands and found that the fractional rms has changed by a factor of $\sim 3$. Since the source exhibited flux drops and intensity-dependent QPO behavior in the course of observation, we further explored the time-dependent behavior of the QPO by extracting the DPS (see Figure \ref{dps_data2}). In the DPS, significant power is observed at the QPO frequency of $\sim 6.6$ Hz during the low flux level, while the QPO power decreases when the intensity increases to higher values, which is consistent with the PDS (see also Figure \ref{pds_seg1_data2}).  

\begin{figure*}

\includegraphics[width=8.9cm,height=7.0cm]{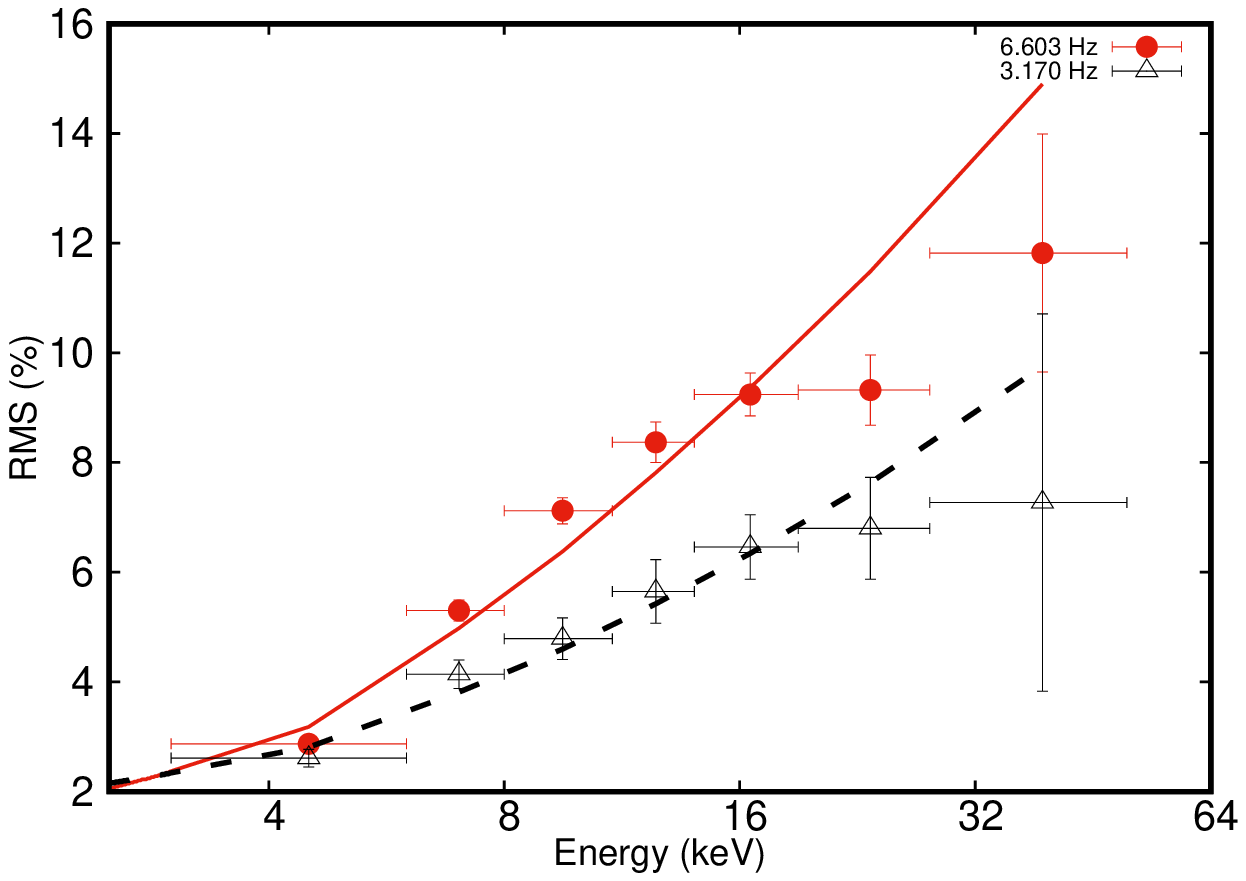}
\includegraphics[width=8.9cm,height=7.0cm]{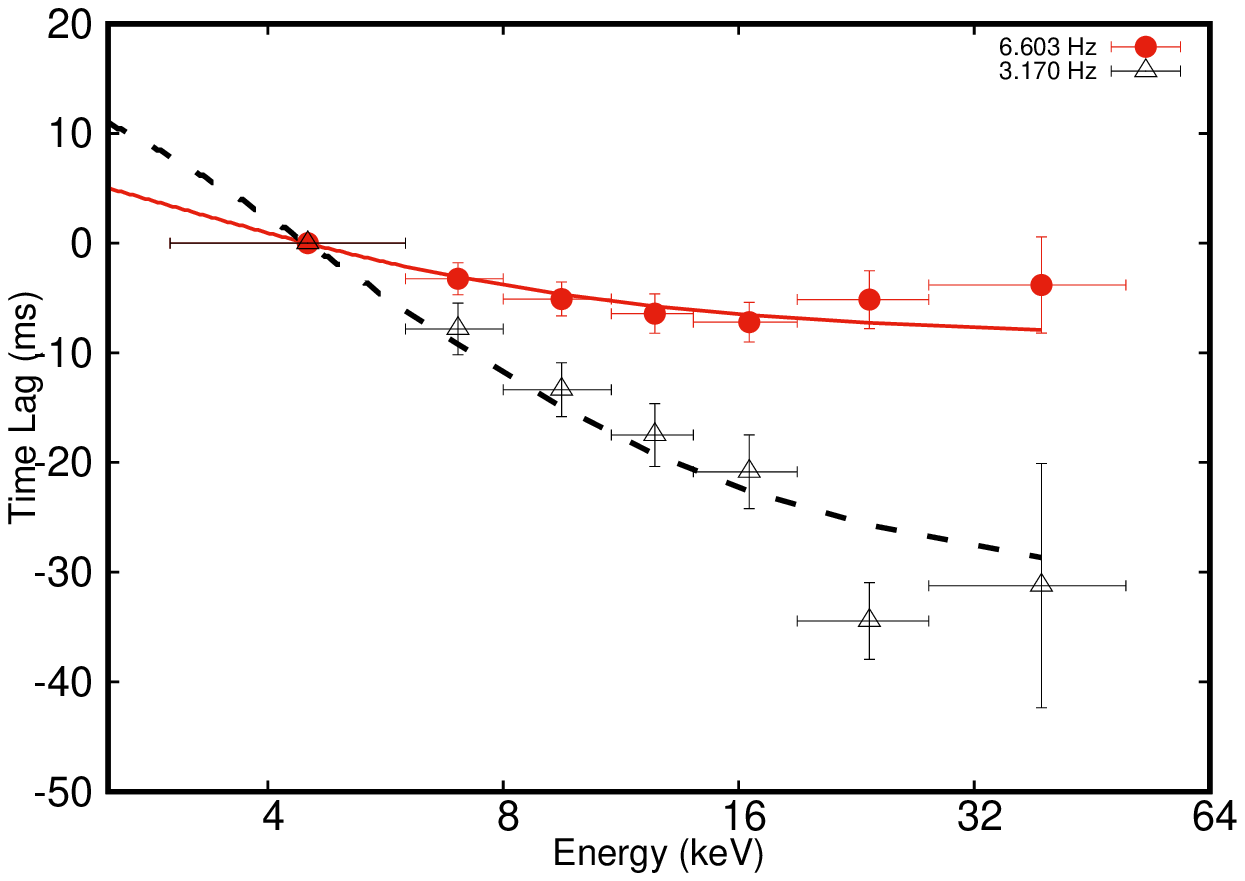}

\caption{Fractional rms (left) and time lag (right) as a function of photon energy from the low intensity level data of second observation. The red filled circle and black open triangle represent the QPO frequency at $\sim 6.60$ Hz and the sub-harmonic at $\sim 3.17$ Hz, respectively. The red solid line and black dotted line represent the model fit derived from the fluctuation propagation model for the QPO and the sub-harmonic, respectively. (see text for details).}  
\label{rms_lag_seg1_data2}
\end{figure*}

It is clear that the QPO is detected only in the low intensity level, thus we estimate the fractional rms and time lag using the data only in the low intensity level. We used different energy bins and computed the fractional rms and time lag at QPO and sub-harmonic frequencies (see Figure \ref{rms_lag_seg1_data2}). The fractional rms at QPO frequency increases from $\sim 2.5$ percent to reach $\sim 12$ percent in the highest energy band of 28--50 keV, while at the sub-harmonic frequency, the fractional rms increases from $\sim 2.5$ to $\sim 7.5$ percent. The time lag at QPO frequency and at the sub-harmonic frequency decreases as the energy increases and the maximum observed time lags are $\sim -7$\,ms and $\sim -35$\,ms for the QPO frequency and the sub-harmonic in the 14--19 keV and 19--28 keV energy bands, respectively. This further implies a soft lag at both QPO ($\sim 6.60$\,Hz) and sub-harmonic ($\sim 3.17$\,Hz) frequencies. 

\subsubsection{Observation 3} 

\begin{figure}
 \includegraphics[width=8.9cm,angle=0]{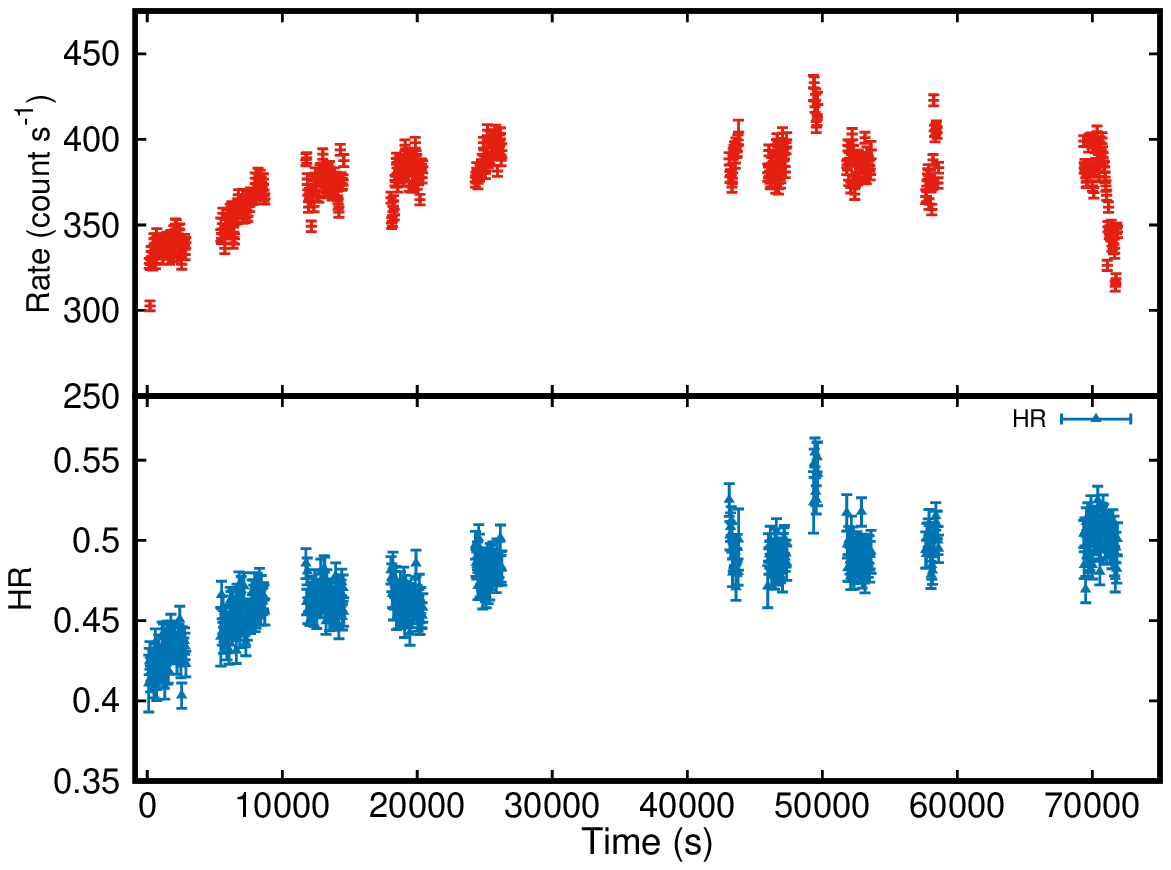}

\caption{The 50\,s binned background subtracted light curve in the 3--50 keV using {\tt LAXPC20} detector (top panel) and HR (bottom panel) from the third observation.} 
\label{lc_data3}
\end{figure}

\begin{figure}

\includegraphics[width=6.2cm,angle=-90]{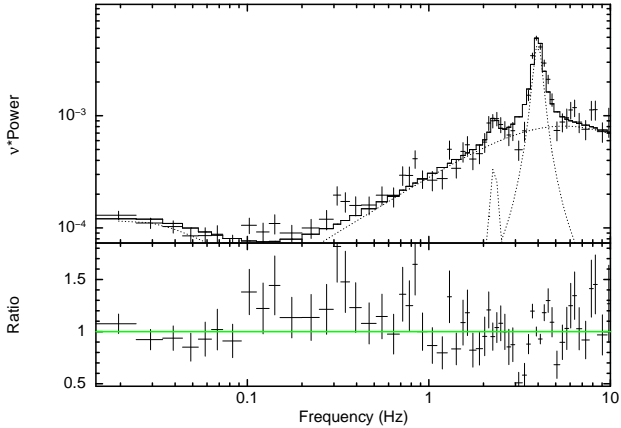}

\caption{The PDS in the 3--50 keV band from the third observation fitted with four Lorentzians. We used {\tt LAXPC20} detector only for this analysis.}
\label{pds_data3}
\end{figure}

\begin{figure*}

\includegraphics[width=8.9cm,height=7.0cm]{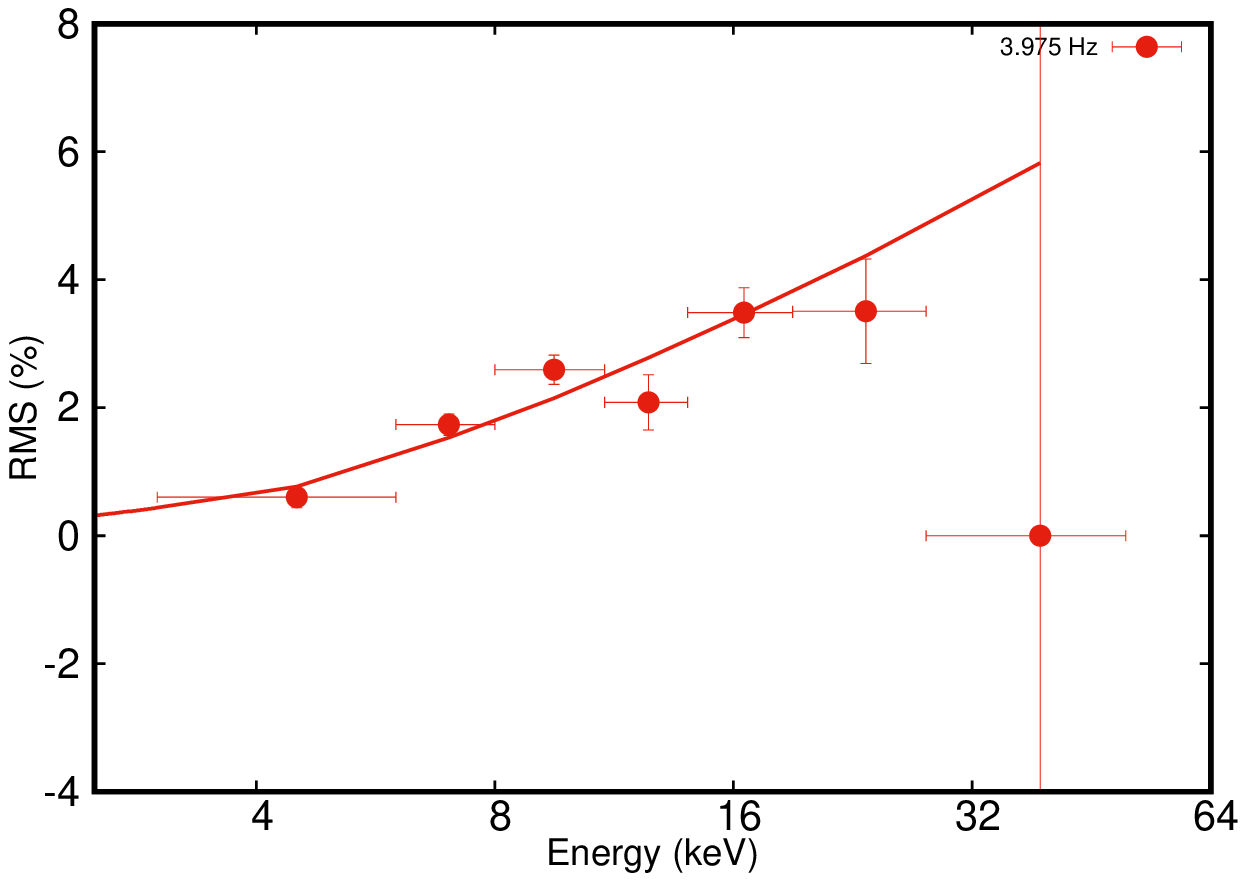}
\includegraphics[width=8.9cm,height=7.0cm]{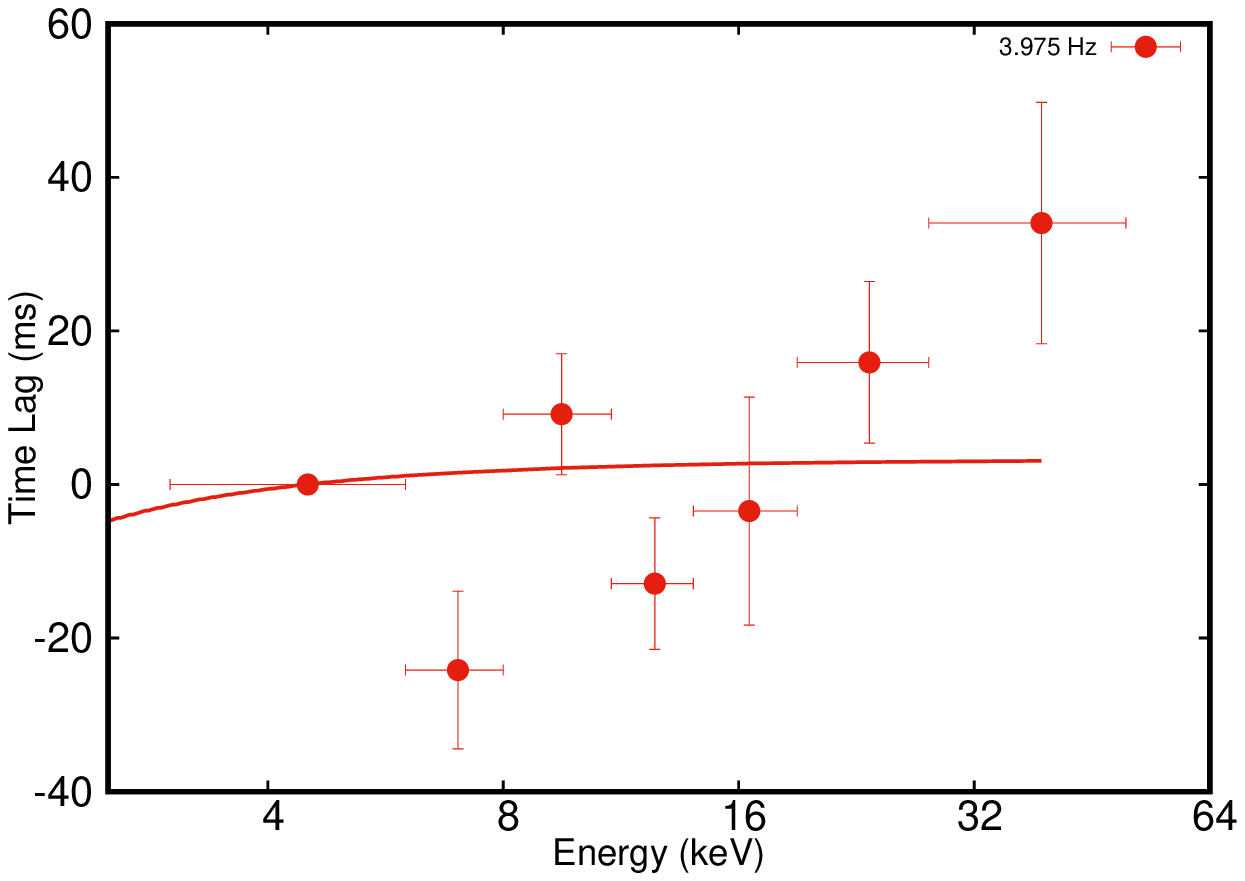}

\caption{Fractional rms (left) and time lag (right) as a function of photon energy from the third observation at the QPO frequency $\sim 3.98$ Hz.}
\label{rms_lag_data3}
\end{figure*}

In the third observation, the {\tt LAXPC30} detector was no longer operational and the {\tt LAXPC10} detector was operating at low gain. Thus, we generated the light curve from {\tt LAXPC20} detector only in the 3--50 keV band with a time bin size of 50\,s, which is shown in the top panel of Figure \ref{lc_data3}. From the plot, it is clear that the source did not significantly vary during the observation and the intensity appears to be constant. Moreover, the calculated HR using the count rates in the 3--7 and 7--16 keV energy bands (see the bottom panel of Figure \ref{lc_data3}) did not show any significant variation. We extracted the PDS in the 0.01--10 Hz using an energy range of 3-50 keV and fitted the PDS with four Lorentzians (see Figure \ref{pds_data3}). In the power spectrum, we detected the QPO and sub-harmonic features at $\sim 3.98$\,Hz and $\sim 2.29$\,Hz, respectively along with some broad features. We have not detected any time-dependent behavior in this QPO, as observed in the first observation. We estimated the fractional rms and time lags at the QPO frequency from different energy bands and plotted them in Figure \ref{rms_lag_data3}. The fractional rms at the QPO frequency marginally increases, from $\sim 1$ to $\sim 4$ percent, with energy. 

\begin{figure}
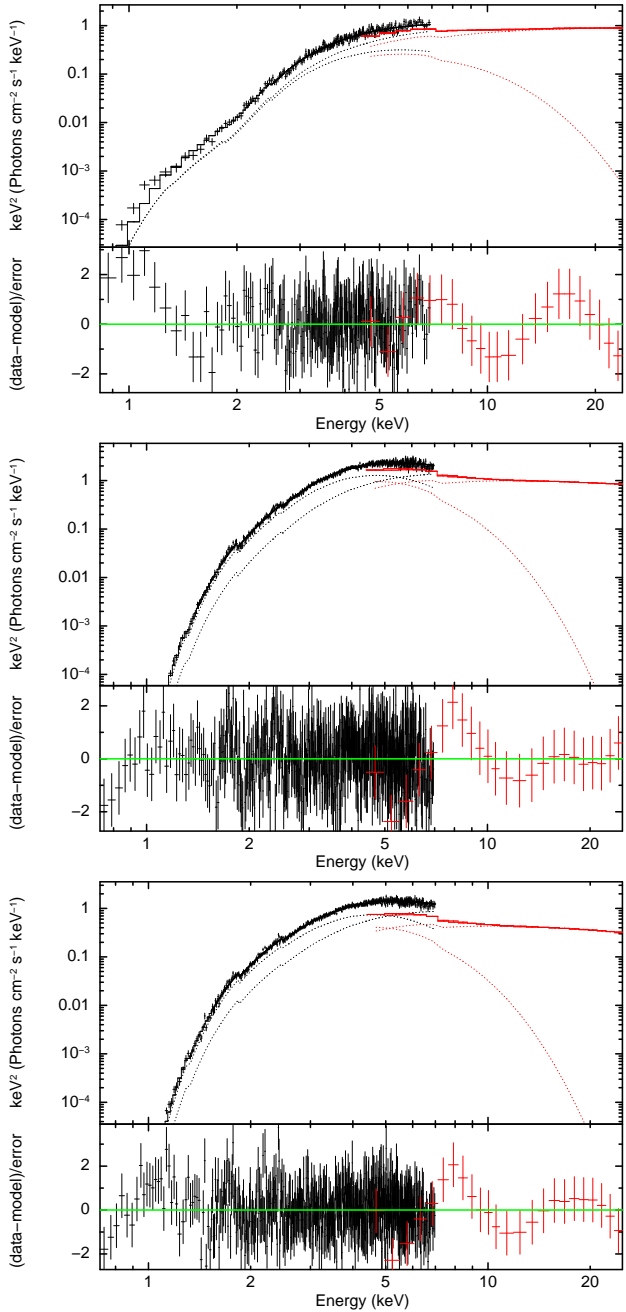


 \includegraphics[width=5.8cm,angle=-90]{f13a.ps}
 \includegraphics[width=5.8cm,angle=-90]{f13b.ps}
 \includegraphics[width=5.8cm,angle=-90]{f13c.ps}

\caption{The 0.7--25 keV broadband X-ray spectrum of \swiftj{} from the first (top), second (middle) and third (bottom) {\it Astrosat} observations. Spectra are fitted with doubly absorbed disk blackbody plus thermal Comptonization model. The black and red data points represent the SXT and {\tt LAXPC20} data, respectively.}

\label{3spec}
\end{figure}

\subsection{Broadband X-ray Spectral Analysis}

We used the broadband X-ray spectral coverage of {\it Astrosat} from SXT and LAXPC instruments, and fitted both spectra of the source simultaneously in the 0.7--25.0 keV energy band. The spectral modeling was performed with {\sc xspec} version 12.9.1p \citep{Arn96}. We used the absorbed multi-color disk blackbody (MCD; {\tt diskbb} in {\sc xspec}) plus thermal Comptonization ({\tt nthcomp}) model components to explain the broadband X-ray spectral features of the source. The absorption is incorporated by Tuebingen-Boulder Inter-Stellar Medium absorption model \citep[{\tt tbabs};][]{Wil00}. The MCD \citep{Mit84} component explains the X-ray emission from the accretion disk and provides the seed photons for the Comptonization. Thus, we kept the MCD temperature and seed photon temperature (in {\tt nthcomp} model) as the same. The {\tt nthcomp} model \citep{Zdz96, Zyc99} describes the thermal Comptonization in an hot corona. The spectral fitting with this model provides a worse fit in all cases with residuals in soft energies. Thus, we added a partial covering fraction absorption ({\tt pcfabs} in {\sc xspec}) model and this model fit resulted in a better fit in all cases with $\Delta \chi^2 = -41.4$ to $-745.0$  for 2 degrees of freedom compared to the absorbed MCD plus thermal Comptonization. In the simultaneous SXT and LAXPC fit, we used a multiplicative constant to address the cross-calibration uncertainties between the instruments. A 3 percent model systematic error is introduced in the model fitting. The quoted parameter uncertainties are at a 90\% confidence level. 

We fitted the observed spectra from the three observations (see Figure \ref{3spec}) and the best-fit model parameters are given in Table \ref{spec_params}. The electron plasma temperature ($\rm kT_{e}$) was not constrained and hence fixed at 70 keV. The first absorption column density ($N_{\rm H tbabs}$) changed from 2.1 to $5.2 \times 10^{22}~\rm cm^{-2}$ between the first and second observations, while the second absorption component exhibits a marginal increasing trend in all three observations. However, the covering fraction and inner disk temperature show a marginal decreasing trend as the time evolves. In addition, the yielded photon index increases from $\sim 1.9$ to $\sim 2.4$ in these observations.

The source showed time-dependent and intensity-dependent QPO behaviors in the first and second observations, respectively. Thus, we extracted the spectra from different time segments to study the spectral characteristics of the source related to these behaviors. The spectral modeling of two spectra from the different time segments in the first observation suggests no clear variations between the best-fit model parameters (see Table \ref{spec_params}). However, in the second observation, the partial covering fraction and inner disk temperature increase between low and high intensity levels, while the change in the power law index is marginal. 

We also computed the unabsorbed flux in the 0.7--25 keV band using the convolution model {\tt cflux}. The flux increases by a factor of $\sim 3$ in the second observation compared to the first, while the flux dropped by a factor of $\sim 2$ in the third observation. It is also noted that the disk fraction increases from $\sim 30\%$ to $\sim 60\%$ in these observations, which suggests a more dominant disk emission in the later observations.   

\begin{table*}
\tablecolumns{7}
\setlength{\tabcolsep}{3.0pt}
\tablewidth{320pt}
	\caption{Broadband X-ray Spectral Parameters for \swiftj{}}
 	\begin{tabular}{@{}ccccccccccccc@{}}
	\hline
	\hline
\colhead{Obs} & \colhead{Seg} & \colhead{$N_{\rm H tbabs}$} & \colhead{$N_{\rm H}$} & \colhead{$f$} & \colhead{$\rm kT_{in}$} & \colhead{$\rm N_{diskbb}$} & \colhead{$\Gamma$} & \colhead{$\rm N_{nthcomp}$} & \colhead{$\rm F_{disk}$} & \colhead{$\rm F_{Total}$} & \colhead{$\rm Ratio$}&\colhead{$\rm \chi^2/ d.o.f$} \\

\hline
O1 & - & $2.06^{+0.99}_{-1.09}$ & $9.33^{+0.55}_{-0.56}$ & $0.968^{+0.018}_{-0.028}$ & $1.74^{+0.22}_{-0.22}$ & $7.60^{+3.95}_{-2.32}$ & $1.87^{+0.09}_{-0.11}$ & $0.16^{+0.05}_{-0.04}$ & $1.38^{+0.34}_{-0.35}$ & $5.09^{+0.25}_{-0.21}$ & 0.27 & 235.8/255 \\

      & 1 & $2.41^{+1.14}_{-1.17}$ & $9.66^{+0.69}_{-0.69}$ & $0.960^{+0.023}_{-0.038}$ & $1.68^{+0.23}_{-0.23}$ & $8.28^{+5.02}_{-2.71}$ & $1.89^{+0.09}_{-0.11}$ & $0.17^{+0.06}_{-0.04}$ & $1.31^{+0.34}_{-0.35}$ & $5.11^{+0.30}_{-0.25}$ & 0.26 & 180.9/187 \\

      & 2 & $3.53^{+2.47}_{-2.39}$ & $9.38^{+1.79}_{-1.18}$ & $0.930^{+0.056}_{-0.160}$ & $1.65^{+0.22}_{-0.22}$ & $11.49^{+8.42}_{-4.13}$ & $1.90^{+0.09}_{-0.11}$ & $0.19^{+0.07}_{-0.05}$ & $1.66^{+0.41}_{-0.40}$ & $5.78^{+0.61}_{-0.43}$ & 0.29 & 109.6/101 \\

O2 & - & $5.20^{+0.23}_{-0.24}$ & $10.72^{+0.87}_{-0.81}$ & $0.829^{+0.017}_{-0.018}$ & $1.10^{+0.06}_{-0.06}$ & $362.16^{+139.57}_{-93.79}$ & $2.25^{+0.06}_{-0.06}$ & $0.61^{+0.12}_{-0.10}$ & $9.73^{+1.06}_{-0.86}$ & $16.96^{+1.50}_{-1.21}$ & 0.57 & 504.5/535 \\

      & 1 & $5.06^{+0.31}_{-0.33}$ & $9.43^{+1.13}_{-1.01}$ & $0.796^{+0.029}_{-0.032}$ & $1.10^{+0.07}_{-0.07}$ & $270.70^{+129.01}_{-80.04}$ & $2.21^{+0.06}_{-0.06}$ & $0.46^{+0.10}_{-0.08}$ & $7.33^{+0.99}_{-0.76}$ & $12.93^{+1.36}_{-1.05}$ & 0.57 & 436.44/462 \\

      & 2 & $4.92^{+0.42}_{-0.42}$ & $10.25^{+0.99}_{-0.86}$ & $0.861^{+0.025}_{-0.029}$ & $1.23^{+0.08}_{-0.08}$ & $251.33^{+105.74}_{-67.61}$ & $2.25^{+0.07}_{-0.07}$ & $0.66^{+0.15}_{-0.12}$ & $10.69^{+1.11}_{-0.91}$ & $19.40^{+1.67}_{-1.31}$ & 0.55 & 408.62/404 \\

O3 & - & $5.06^{+0.18}_{-0.18}$ & $12.16^{+1.07}_{-1.01}$ & $0.786^{+0.022}_{-0.023}$ & $1.04^{+0.07}_{-0.07}$ & $287.14^{+138.44}_{-85.35}$ & $2.40^{+0.06}_{-0.07}$ & $0.46^{+0.10}_{-0.08}$ & $6.19^{+0.89}_{-0.69}$ & $10.97^{+1.26}_{-0.98}$ & 0.56 & 498.45/537 \\

\hline
\end{tabular} 
\tablecomments {(1) Observation; (2) segment used; (3) neutral hydrogen column density in units of $10^{22}~\rm cm^{-2}$ from {\tt tbabs} model; (4) neutral hydrogen column density in units of $10^{22}~\rm cm^{-2}$ from {\tt pcfabs} model; (5) covering fraction; (6) inner disk temperature in keV  (7) {\tt diskbb} normalization ; (8) photon index; (9) {\tt nthcomp} normalization; (10)-(11) the unabsorbed disk flux and total flux in units of $10^{-9}\rm~erg~cm^{-2}~s^{-1}$ in the 0.7--25 keV band derived using {\tt cflux}; (12) Ratio of the disk flux to the total flux; (13) $\chi^2$ statistics and degrees of freedom.}
\label{spec_params}
\end{table*}

\subsection{Fitting the energy dependence timing properties}

The motivation here is to quantify the energy-dependent fractional rms and time lags of \swiftj{} as measured by LAXPC. In order to do so, we invoke a single zone stochastic propagation model, which is characterized by only three parameters: the normalized variations of the electron temperature of the inner hot flow ($\delta T_{\rm e}$), inner disk temperature ($\delta T_{\rm in}$) and the phase angle between them ($\phi=2\pi f \tau_{D}$, where $f$ is the QPO frequency and $\tau_{D}$ is the time lag). However, for model fitting we use the three parameters, $\delta T_{\rm e}$, $\tau_{D}$ and the attenuation parameter, which quantifies the attenuation of propagations from the corona to disk, given by the ratio $\delta T_{\rm in}/\delta T_{\rm e}$. A detailed description of the model is presented in \citet{Maq19}. The geometry of the system is assumed to be that of a truncated standard disk with a hot inner flow and \citet{Maq19} applied the model to the hard state of Cygnus X--1. While their analysis was for the broadband noise, here we apply the same model for the QPOs observed. Thus, in contrast to the general idea of fluctuation propagating inwards, we assume that an oscillation originates in the inner hot region causing variations in the electron temperature, which then propagates outwards, reaches the outer truncated disk causing variations in the inner disk temperature, but after a time delay.  

We first fitted the fractional rms and time lag at the QPO frequency and obtained the fit parameters as given in Table \ref{modelfit}. For each of the 2nd harmonic frequencies, we successfully fitted the model by fixing the phase parameter at twice the value of the phase of the respective QPO. The model fitting shows similar time lag for the QPO frequency of 1.56 Hz and 2nd harmonic of 3.12 Hz in the first observation. This trend was also seen for the QPO and 2nd harmonic frequencies at 1.70 Hz and 3.48 Hz, respectively. However, in the case of sub-harmonic frequency (3.17 Hz), we were not able to fit the observed time lag by fixing the phase value at half of the phase of the QPO at 6.6 Hz. However, keeping phase as a free parameter, we were able to fit the time lag for sub-harmonic frequency. Thus, the model suggests that the fitting of time lag for the sub-harmonic frequency requires equal phase change rather than equal time delay. In the case of QPO frequency of 3.97 Hz, we were able to constrain the fractional rms and time lag with a systematic of 17\% suggesting that perhaps the temporal behavior is more complex than the simple model assumed here. 

\begin{table}
\tabletypesize{\small}
\tablecolumns{5}
\setlength{\tabcolsep}{4.0pt}
\tablewidth{220pt}
	\caption{Best-fit Parameters from the Fluctuation Propagation Model}
 	\begin{tabular}{@{}ccccc@{}}
	\hline
	\hline
\colhead{Frequency} & \colhead{$\delta T_{\rm e}$} & \colhead{$\tau_{D}$} & \colhead{$\delta T_{\rm in}/\delta T_{\rm e}$} & \colhead{$\rm \chi^2/ d.o.f$} \\
     	 (Hz)      &		 & 	(ms)	  &  &  \\
\hline
1.561 & $0.29^{+0.03}_{-0.02}$ & $-1.49^{+3.65}_{-3.66}$ & $0.087^{+0.006}_{-0.006}$ & 25.0/18 \\
3.127 & $2.27^{+0.00}_{-1.88}$ & $-1.49(f)$ & $0.004^{+0.000}_{-0.004}$ & 7.5/19 \\
 
1.740 & $0.29^{+0.03}_{-0.03}$ & $-3.49^{+4.28}_{-4.28}$ & $0.090^{+0.006}_{-0.006}$ & 28.1/18 \\
3.489 & $1.19^{+0.00}_{-0.75}$ & $-3.49(f)$ & $0.008^{+0.011}_{-0.008}$ & 14.1/19 \\

6.603 & $0.09^{+0.02}_{-0.02}$ & $-12.74^{+3.08}_{-3.12}$ & $0.061^{+0.004}_{-0.004}$ & 17.9/10 \\
3.170 & $0.15^{+0.03}_{-0.03}$ & $-37.98^{+4.15}_{-4.14}$ & $0.041^{+0.004}_{-0.005}$ & 12.0/10 \\

3.975 & $0.03^{+0.08}_{-0.00}$ & $10.25^{+0.0}_{-0.0}$ & $0.021^{+0.005}_{-0.004}$ & 19.9/10 \\

\hline
\end{tabular} 
\tablecomments {(1) Frequency in Hz; (2) variation in the electron temperature; (3) time lag in ms; (4) ratio of the variation in the inner disk temperature to the variation of electron temperature; (5) $\chi^2$ statistics and degrees of freedom.}
\label{modelfit}
\end{table}

\section{Summary and Discussion}
\label{sec:discu}

We have presented the spectral and timing analysis of the new black hole binary candidate \swiftj{} using {\it Astrosat} observations. The main results of the study are as follows.

\begin{enumerate}
\item We detected QPOs at frequencies of $\sim 1.7$, $\sim 6.6$ and $\sim 4.0$\,Hz.

\item The QPO detected at $\sim 1.56$ Hz drifts to a higher centroid frequency of $\sim 1.74$ Hz in the course of the first observation. 

\item We observed a peculiar behavior in the second observation, where the QPO detected at $\sim 6.6$\,Hz becomes weak or disappears at high intensity levels.

\item The fractional rms of the QPOs increases with photon energy and soft time lag of the order of $\sim 35$ milliseconds were detected for the QPO at 6.6 Hz and its sub-harmonic. 

\item The doubly absorbed disk plus thermal Comptonization model successfully explains the observed broadband spectra and the best-fit spectral parameters varies between the observations.

\item We quantified the frequency-dependent fractional rms and time lag using a single zone stochastic propagation model. Variation of the temperature of the corona and the disk with a time lag between them can explain the energy-dependent temporal behavior. 

\end{enumerate}

Type-C QPOs are mainly detected in the HIMS of BHXRBs \citep{Wij99, Cas05, Mot11}. Their centroid frequencies have a range of 0.1--15 Hz with $Q$ factor ($Q=\nu/\rm FWHM$) of 7--12 and a typical rms of 3--16\% \citep{Cas04, Cas05}. The detected LFQPOs at $\sim 1.7$, $\sim 6.6$ and $\sim 4.0$\,Hz from the {\it Astrosat} observations are consistent with the above-mentioned values. Thus, these LFQPOs are most likely belong to the type-C QPOs \citep{Cas04, Cas05, Mot15}. The QPO detected at 1.56\,Hz drifts to a slightly higher centroid frequency in the course of observation (see Figure \ref{dps_data1}). Similar behavior has been reported in the hard state of the source using {\it NuSTAR} observation, where the QPO frequency drifts from 0.14 to 0.20\,Hz \citep{Xu18}. In addition, a transient QPO at 6 -- 7 Hz has been reported for the source from the {\it NuSTAR} and {\it XMM-Newton} observations in the intermediate state, where the QPO appeared only in the low flux level \citep[see Figure 2 and 4 of][]{Xu19}. We have observed similar behavior in the {\it Astrosat} observation, where the QPO at $\sim 6.6$\,Hz is strongly detected in the low intensity level, while the QPO becomes weak when the source exhibits flaring (see also Figure \ref{lc_data2}).

Rapid flux transitions have been detected in some black hole X-ray binaries in their very high states and such behavior has been proposed to have a jet origin \citep[e.g.,][]{Miy91, Tak97, Cas04, Bel05}. Moreover, these flux transitions are associated with the fast transition between the different types of QPOs. However, the observed scenario in the case of \swiftj{} is phenomenologically distinct compared to other sources, where the QPO becomes weak or is not detected during the hard flaring. The disappearance of QPO associated with rapid flux transition has never been observed in other BHXRBs. This confirms the results from the {\it NuSTAR}, {\it Swift} and {\it XMM-Newton} observations, where they observed a transient QPO in the low flux state alone \citep{Xu19}. Thus, the origin of the rapid changes in the observed properties may be due to thermal and viscous instabilities \citep{Sha73} in accretion disk as has been reported earlier for this source \citep{Xu19}.

For the first time we have studied the fractional rms and time lag as a function of photon energy up to 50 keV for \swiftj{} (see Figure \ref{data1_seg1_2_rm_lag}, \ref{rms_lag_seg1_data2} and \ref{rms_lag_data3}). It is noted that the fractional rms at the QPO and the sub-harmonic frequencies increase with photon energy, while at the 2nd harmonic frequencies the rms seems to be constant. The observed energy dependence of the fractional rms at the QPO frequencies clearly suggests that these LFQPOs are associated with the high-energy component, i.e, the corona. In addition, we have observed soft time lag at QPO and sub-harmonic frequencies. However, at the 2nd harmonic frequencies there is weak/zero time lag. 

In order to quantify the frequency-dependent fractional rms and time lags, we invoke a single zone stochastic propagation model. The model fitting provides physical measures of the time the fluctuation takes to travel from the inner hot region to the outer truncated disk and successfully explains the time lags observed in the source. We observe a similar time lag for the QPO (1.56 and 1.74 Hz) and 2nd harmonic (3.17 and 3.49 Hz) frequencies, while the model fitting provides different time lags for QPO and sub-harmonic frequencies at 6.6 and 3.17 Hz, respectively. Thus, the model suggests that the fitting of time lag for the sub-harmonic frequency requires equal phase change rather than equal time delay.

It is interesting to compare the energy dependent timing properties of \swiftj{} with Cygnus X-1. For a QPO frequency of 1.56 Hz, the time lag between the $\delta T_{\rm in}$ and $\delta T_{\rm e}$ is about $-1.5$ ms, implying that the variation $\delta T_{\rm e}$ occurs first. For Cygnus X-1, the time lag is similar for the same frequency, but is positive, implying that the variation $\delta T_{\rm in}$ occurs first \citep{Maq19}. Thus, while for Cygnus X-1, the corona responds to a stochastic perturbation that occurs in the disk, for \swiftj{} it is opposite, such that the QPO originates in the corona and the fluctuation propagates outwards towards the disk. Nevertheless, the time lags for both systems are of the same magnitude implying that the speed of the fluctuations and the distance traveled are of the same order. For \swiftj{}, the primary driver $\delta T_{\rm e}$ induces a small fluctuation $\delta T_{\rm in}$ and thus the attenuation ratio is small, $\delta T_{\rm in}/\delta T_{\rm e} \sim 0.09$. On the other hand, for Cygnus X-1, since the driver is $\delta T_{\rm in}$, the ratio turns out to be an order of magnitude larger \citep{Maq19}.

The hard intermediate state of BHXRBs is characterized by the decrease in the HR in the hardness-intensity diagram along with the soft spectra, where the index increases up to $\sim 2.5$, and the presence of disk component. In addition, the total fractional rms decreases with softening and the type-C QPOs are widely detected in this state \citep{Bel10}. During the {\it Astrosat} observations, we observed a drop in the HR by a factor of $\sim 2$, the photon index significantly changed from $\sim 1.9$ to $\sim 2.4$, the disk fraction increases steadily to $\sim 60$\% and type-C QPOs are detected. Thus, based on the timing, hardness ratio, and spectral analysis, we have identified the source to be mainly in the HIMS of BHXRBs. 

The column density varies between the observations and the inferred values are generally consistent with previous studies with {\it Swift} and {\it NuSTAR} observations \citep{Xu18}. Thus, our analysis with {\it Astrosat} confirms the highly absorbed nature of \swiftj{}. The inner disk temperature increased by $\sim 12\%$ when the flux increased during the second observation, which is consistent with earlier studies by \citet{Xu19}, where they observe a decrease in disk temperature by $\sim 15\%$ when the flux dropped. The observations conducted with {\it Swift} and {\it NuSTAR} on 2018 February 16 identified the source in the hard state with an index of $\sim 1.63$ \citep{Xu18}, while the {\it Astrosat} observation conducted on 2018 February 20-21 identified the source in the hard-intermediate state with an index of $\sim 1.9$. The joint {\it XMM-Newton} and {\it NuSTAR} observations conducted on 2018 February 25 identified the source to be in the hard-intermediate state with an index of $\sim 2.1$. Thus, it seems that the observed spectral evolution of \swiftj{} is similar to the other Galactic BHXRBs \citep{Bel10, Bel11}.    

Previous studies of \swiftj{} with {\it Swift} and {\it NuSTAR} observations found a strong reflection component in the bright hard state \citep{Xu18}. In the following observations after $\sim 8$ days, the source was identified to be in the intermediate state of BHXRBs, where the reflection feature becomes weak. One possible explanation for the observed scenario is related to the truncation of accretion disk, where the disk reached the innermost stable circular orbit (ISCO) in the bright hard state and a strong reflection component is detected, while the accretion disk receded from the compact object in the intermediate state which leads to the detection of weak reflection component \citep{Xu19}. However, such reflection component is not apparent in the {\it Astrosat} observations, where the source is mostly identified in the intermediate state. Thus, the non-detection of reflection features in the {\it Astrosat} spectra may be because of the intrinsic weakness of the reflection component in the intermediate state. Such a weak component may be hard to detect with the current uncertainties in the response matrix of {\it Astrosat}.

\section*{Acknowledgements}

We thank the anonymous referee for the constructive comments and suggestions that improved this manuscript. This work was started in the IUCAA-APT Mini School on X-ray Astronomy Data Analysis held at Providence Women's College, Kozhikode and we thank the organizers for the hospitality. ART and GM acknowledge the IUCAA Visitors Program. The research is based on the results obtained from the {\it Astrosat} mission of the Indian Space Research Organization (ISRO), archived at the Indian Space Science Data Centre (ISSDC). This work has used the data from the LAXPC and SXT instruments. We thank the LAXPC Payload Operation Center (POC) and the SXT POC at TIFR, Mumbai for providing the data via the ISSDC data archive and the necessary software tools. \\

{\it Facility}: {\it Astrosat}, ADS, HEASARC \\  
{\it Software}: LaxpcSoft, SXT Software, HEASOFT (v 6.22), XSPEC \citep[v 12.9.1p;][]{Arn96}


\end{document}